# Incoherent non-Hermitian skin effect in photonic quantum walks

Stefano Longhi[1,2,*]

[1]*Dipartimento di Fisica, Politecnico di Milano, Piazza L. da Vinci 32, I-20133 Milano, Italy.*

[2]*IFISC (UIB-CSIC), Instituto de Fisica Interdisciplinar y Sistemas Complejos, E-07122 Palma de Mallorca, Spain.*

[*]Corresponding author: S.L. (email: stefano.longhi@polimi.it)

## Abstract

**The non-Hermitian skin effect describes the concentration of an extensive number of eigenstates near the boundaries of certain dissipative systems. This phenomenon has raised a huge interest in different areas of physics, including photonics, deeply expanding our understanding of non-Hermitian systems and opening up new avenues in both fundamental and applied aspects of topological phenomena. The skin effect has been associated to a nontrivial point-gap spectral topology and has been experimentally demonstrated in a variety of synthetic matter systems, including photonic lattices. In most of physical models exhibiting the non-Hermitian skin effect full or partial wave coherence is generally assumed. Here we push the concept of skin effect into the fully incoherent regime and show that rather generally (but not universally) the non-Hermitian skin effect persists under dephasing dynamics. The results are illustrated by considering incoherent light dynamics in non-Hermitian photonic quantum walks.**

# Introduction

The physics of dissipative classical and quantum systems has received a renewed and growing attention recently [1-8], providing a flourishing and impactful area of research with the prediction and observation of a wealth of unprecedented physical phenomena propitious for future applications, especially in photonics [2-7]. Recent progress in the field of non-Hermitian (NH) topological phases has shown great promise in discovering new types of topological phenomena beyond the Hermitian paradigm. A representative example is the NH skin effect [9-11], which has opened new avenues for understanding the elusive NH physics [9-37]. The phenomenon, unique to the NH band theory [9,11,12], describes the strong dependence of the energy spectrum of certain NH systems on boundary conditions and the exotic property for which an extensive number of eigenstates reside at the boundaries of a system rather than being uniformly distributed throughout the bulk. Featuring the breakdown of conventional bulk-boundary correspondence [9,13], the NH skin effect was early introduced in one-dimensional (1D) systems [9,10] and was shown to be related to a non-trivial point-gap spectral topology [15,16], with a characteristic fingerprint of a persistent directional current in the bulk dynamics [14,16]; for recent reviews see [29,31,32,35-37]. Extensions of the NH skin effect to two (2D) and higher dimensional systems have been investigated [23,24,28,30], suggesting that in high dimensions the skin effect is a universal phenomenon [30]. The NH skin effect and failure of the conventional bulk-boundary correspondence have been experimentally observed in a variety of synthetic models of NH matter [18-21,25-28,34], including photonic lattices [19,21,26,27], and its persistence in fully quantum-mechanical models of open quantum systems has been pointed out [38-41]. In most models displaying the NH skin effect so far considered full or partial wave coherence is usually assumed, as in effective NH Hamiltonian or Lindblad master equation models, however the fate of the NH skin effect in the fully incoherent regime, where quantum coherence is lost and the dynamics behaves fully classically [42-46], remains largely unexplored. Incoherent or partial coherent hopping dynamics is commonplace in many complex physical, chemical and biological systems out of equilibrium [45,47,48], and it is thus of main relevance to extend the idea of skin effect to incoherent models.

In this work we consider the fully incoherent regime of NH models and show that rather generally (but not universally) the NH skin effect persists under incoherent (dephasing) dynamics. The results are illustrated by considering incoherent photonic quantum walks in synthetic mesh lattices [19,21,26,43,49,50], which should provide an experimentally accessible platform for the observation of incoherent NH skin effect in bulk dynamics.

## Results

**Incoherent non-Hermitian skin effect.**

Let us consider a rather arbitrary NH lattice or network system, typically a 1D or 2D system, with open boundaries of given shape, and let us indicate by $\hat{H}$ the tight-binding Hamiltonian of the system under fully coherent wave dynamics, which in physical space is described by a NH matrix $H_{n,m} = \langle n|\hat{H}m\rangle$, where $\{|n\rangle\}$ describes the Wannier basis, $n=1,2,3,\ldots,N$ labels the sites (or nodes) of the lattice, and $N$ is the total number of sites (Fig.1**a**). In 1D systems, the NH skin effect arises rather generally in the presence of non-reciprocal hopping amplitudes and the fingerprint is a non-trivial point gap topology of the periodic-boundary energy spectrum [15,16,33,36]. In 2D systems the NH skin effect is a rather universal phenomenon related to a non-vanishing spectral area on the complex plane covered by the periodic-boundary spectrum [30]. Under fully coherent dynamics, the wave function of the system evolves according to $|\psi(t)\rangle = \hat{U}_{coh}(t)|\psi(0)\rangle$, possibly with normalization of the wave function norm at each time instant, where $\hat{U}_{coh}(t) = \exp(-i\hat{H}t)$ is the coherent propagator. The incoherent dynamics is obtained as a dephasing process by assuming that, at the time instants $t_\alpha = \Delta t, 2\Delta t, \ldots, \alpha\Delta t, \ldots$ spaced by the time interval $\Delta t$, the phase of the wave function amplitude $\psi_n(t_\alpha) = \langle n|\psi(t_\alpha)\rangle$ is randomized, i.e. it is multiplied by a random phase $\phi_n^{(\alpha)}$, with $\phi_n^{(\alpha)}$ uncorrelated in both site index $n$ and time step $\alpha$. After letting $\rho_{n,m}(t) = \overline{\psi_n^*(t)\psi_m(t)}$, where the overbar denotes statistical average over the random phase distribution, after each time $t_\alpha$ one clearly has $\rho_{n,m} = 0$ for $n \neq m$, i.e. dephasing drives the dynamics into the classical regime (Fig.1**b**), which is fully described by a discrete-time map for the (unnormalized) occupation probabilities $P_n(t) = \rho_{n,n}(t)$ of the various lattice sites (classical random walk). Assuming a short time interval between successive stochastic phases, such that $\hat{U}_{coh}(\Delta t) = \exp(-iH\Delta t)$ can be expanded up to second order in $\Delta t$ as $\hat{U}_{coh}(\Delta t) \simeq 1 - iH\Delta t - (1/2)H^2\Delta t^2$, the average occupation probabilities satisfy the master equation (see Sec.S1 of the Supplementary Material for technical details)

$$\frac{dP_n}{dt} = \sum_{l=1}^{N} M_{n,l} P_l(t). \tag{1}$$

In the above equation, $M_{n,l}$ are the elements of the Markov transition matrix of the random walk, given by

$$M_{n,l} = 2\delta_{n,l}\text{Im}(H_{n,n}) + \Delta t \left[ |H_{n,l}|^2 - \delta_{n,l}\text{Re}\left\{\sum_q H_{n,q} H_{q,n}\right\} \right] \tag{2}$$

The dynamics becomes trivially decoupled when $\text{Im}(H_{n,n}) \neq 0$, i.e. when non-Hermiticity in the system is introduced by on-site gain and/or loss terms, since in this case at leading order we can disregard the term of order $\sim \Delta t$ in Eq.(2) and the Markov transition matrix $M$ is diagonal. In the following, we will therefore focus our attention to the most interesting case where $H_{n,n}$ is real and non-Hermiticity in the system is introduced by violation of the Hermitian conjugation relation $H_{n,m} = H_{m,n}^*$ for some $n \neq m$. In this case one obtains

$$M_{n,l} = \Delta t \left[ |H_{n,l}|^2 - \delta_{n,l}\text{Re}\left\{\sum_q H_{n,q} H_{q,n}\right\} \right] \tag{3}$$

Note that the incoherent dynamics described by the master equation (1) with Markov transition matrix $M$ can be formally viewed as the coherent dynamics of an associated Hamiltonian system with Wick-rotated matrix Hamiltonian $H'=iM$. For $n \neq l$, $M_{n,l} = \Delta t |H_{n,l}|^2$ represents the (incoherent) hopping rate from site $l$ to site $n$ in the classicalized dynamical regime. In the Hermitian limit, $H_{n,l} = H_{l,n}^*$, the total probability $\sum_{n=1}^{N} P_n(t) = 1$ is conserved, all eigenvalues $\lambda_l$ of $M$ are real with $\lambda_l \leq 0$, and there is one zero eigenvalue $\lambda_1 = 0$ of the most dominant eigenstate, corresponding to the uniform distribution eigenvector $P_n = 1/N$. In this case the incoherent dynamics basically drives any initial state into the steady-state corresponding to equal probability of excitation in each site of the lattice, regardless of the shape of the boundaries. In the non-Hermitian case, i.e. when $H_{n,l} \neq H_{l,n}^*$ for some $l \neq n$, the total probability is not conserved, the eigenvalues can be real or appear in complex conjugate pairs, and the constraint $\text{Re}(\lambda_l) \leq 0$ can be violated. However, at each time step we can renormalize the probabilities by letting $P_n(t) \to P_n(t)/\sum_l P_l(t)$, so that probability conservation is restored and under OBC the system is driven toward the stationary state corresponding to the eigenvector of *M* with the largest real part of corresponding eigenvalue. Like for the coherent Hamiltonian *H*, we say that the system displays incoherent NH skin effect whenever in the large *N* limit an extensive number of eigenstates of the Markov transition matrix *M*, or equivalently of *H'*, are localized at the boundaries or corners of the lattice.

The appearance of the NH skin effect is generally (but not exclusively) related to non-reciprocal couplings [36], i.e. $|H_{n,m}| \neq |H_{m,n}|$ for some $n \neq m$, which is a frequent source of non-Hermiticity in out of equilibrium systems. In this case one speaks about *non-reciprocal skin effect*. Since $|H_{n,m}| \neq |H_{m,n}|$ implies $M_{n,m} \neq M_{m,n}$, the system is likely to display non-reciprocal skin effect under incoherent dynamics as well. However, the NH skin effect can appear also in models with reciprocal hopping, i.e. $|H_{n,m}| = |H_{m,n}|$ for any $n \neq m$. This kind of edge localization is dubbed the *reciprocal skin effect* [20,30,36]. Since the elements of *M* are insensitive to the phases of the elements of *H*, the reciprocal NH skin effect in coherent models is expected to be washed out under incoherent (dephasing) dynamics. This means that, as a general rule of thumb, the skin effect persists under incoherent dynamics when it originates from non-reciprocal couplings. Such general results can be exemplified by considering two significant NH models displaying either non-reciprocal or reciprocal NH skin effects.

*Non-reciprocal skin effect.*

The first example, which provides a paradigmatic and simplest model displaying the NH skin effect based on non-reciprocal hopping amplitudes, is the clean Hatano-Nelson model [51,52] in a 1D lattice (Fig.2**a**). The Bloch Hamiltonian of this model is given by $H(k) = \kappa_1 \exp(ik) + \kappa_2 \exp(-ik)$, where $\kappa_1, \kappa_2$ are the non-reciprocal left/right hopping rates. Correspondingly, in Bloch space the Markov matrix describing the incoherent Hatano-Nelson model has the form $M(k) = \Delta t[-2\kappa_1\kappa_2 + \kappa_1^2 \exp(ik) + \kappa_2^2 \exp(-ik)]$. The eigenvalues and corresponding eigenvectors of both matrices *H* and *M* are strongly dependent on the boundary conditions, i.e. periodic (PBC) or open (OBC) boundaries, as illustrated in Fig.2**b**. In the coherent regime, the eigenenergies of *H*(*k*) under PBC describe an ellipse in complex energy plane of equation $E = H(k) = \kappa_1 \exp(ik) + \kappa_2 \exp(-ik)$, where *k* varies in the Brillouin zone $-\pi \leq k < \pi$. Under OBC the spectrum is entirely real and covers the interval $(-2\sqrt{\kappa_1\kappa_2}, 2\sqrt{\kappa_1\kappa_2})$; it is obtained from the relation *E*=*H*(*k*) by complexification of the Bloch wave number $k$, which should vary on the generalized Brillouin zone [9,12] $k = q - i\gamma$, with $-\pi \leq q < \pi$ and $\gamma = (1/2)\log(\kappa_2/\kappa_1)$. Likewise, in the incoherent regime the eigenvalues of $M(k)$ under PBC describe an ellipse in complex plane of equation $E=M(k)$, with $-\pi \leq k < \pi$, whereas under OBC the spectrum is entirely real and describes the interval $(-4\kappa_1\kappa_2\Delta t, 0)$. Both coherent and incoherent skin effects are clearly demonstrated by the exponential localization of the eigenvectors of $\widehat{H}$ and $\widehat{M}$ at the edges of the lattice under OBC, which arises from the complexification of $k$. In bulk dynamics, the fingerprint of the non-reciprocal skin effect is an asymptotic drift of excitation along the lattice, i.e. a persistence current in the system, regardless of the initial state of the system [14,16]. For the coherent Hatano-Nelson model, the drift velocity is given by [14] $v_{coh} = (\kappa_1 + \kappa_2)$. Likewise, in the incoherent version of the Hatano-Nelson model an asymptotic drift of the probability distribution $P_n(t)$ is observed in the bulk at the drift velocity

$$v_{inc} = \Delta t(\kappa_2^2 - \kappa_1^2). \tag{4}$$

In fact, for an infinitely-extended lattice the solution to the master equation (1) for the incoherent Hatano-Nelson model can be given in terms of the integral representation

$$P_n(t) = \int_{-\pi}^{\pi} dk\, F(k) \exp[ikn + \lambda(k)t] \tag{5}$$

where $\lambda(k) = M(k) = \Delta t \kappa_1^2 \exp(ik) + \Delta t \kappa_2^2 \exp(-ik) - 2\Delta t \kappa_1 \kappa_2$ is the PBC spectrum of $M$ and the spectral amplitude $F(k)$ entering in Eq.(5) is determined by the initial probability distribution $P_n(0)$. For example, if at initial time the system is prepared at site $n=0$, i.e. $P_n(0) = \delta_{n,0}$, one has $F(k) = 1/(2\pi)$. In the long time limit, the integral on the right hand side of Eq.(5) is dominated by the spectral contribution at around $k = k_0 = 0$, where the real part of $\lambda(k)$ takes its largest value, and can be calculated using standard asymptotic methods, yielding

$$P_n(t) \sim F(k_0) \sqrt{\frac{2\pi}{\Delta t(\kappa_1^2 + \kappa_2^2)t}} \exp[\Delta t(\kappa_2 - \kappa_1)^2 t]$$

$$\times \exp\left\{-\frac{(n - v_{inc}t)^2}{2\Delta t(\kappa_1^2 + \kappa_2^2)t}\right\} \tag{6}$$

where $v_{inc}$ is given by Eq.(4). Equation (6) clearly shows that the bulk dynamics of the probability distribution in the lattice drifts at the speed $v_{inc}$ and spreads around its center of mass diffusively, i.e. the width of the distribution increases in time as $\sqrt{t}$. An example of bulk dynamics for the Hatano-Nelson model both in the coherent and incoherent regimes, corresponding to initial excitation of site $n=0$ of the lattice and displaying a drift motion, is shown in Fig.2**c**. We mention that the non-reciprocal skin effect under incoherent dynamics is observable in other models, such as in the 1D Su-Schrieffer-Heeger NH model [9] or rather generally in non-reciprocal 2D models, such as those displaying corner states [30,36], i.e. in which all eigenstates are localized at corners of the system.

*Reciprocal skin effect.*

As a second example, let us consider a 2D square lattice with reciprocal hopping amplitudes, namely with an Hermitian coupling $\kappa_1$ in the horizontal direction and an anti-Hermitian hopping amplitude $i\kappa_2$ in the vertical direction (Fig.3**a**). The Bloch Hamiltonian of the system reads $H(k_x, k_y) = 2\kappa_1 \cos k_x + 2i\kappa_2 \cos k_y$. This model shows the dubbed geometric reciprocal NH skin effect [30]: the spectrum of $H(k)$ under PBC covers a non-vanishing area in complex plane, which implies the appearance of skin modes for rather arbitrary system shape, except for square geometries [30]. The corresponding form of the Markov transition matrix *M* in Bloch space is readily calculated from Eq.(3) and reads explicitly $M(k_x, k_y) = 2\Delta t(\kappa_2^2 - \kappa_1^2) + 2\Delta t(\kappa_1^2 \cos k_x + \kappa_2^2 \cos k_y)$. Note that, while *H(k)* is a complex function and covers a non-vanishing area (a square) in complex energy plane, *M(k)* is entirely real and describes a segment on the real axis, as shown in Fig.3**b**. Therefore, while the system under coherent dynamics displays the reciprocal skin effect for rather arbitrary shape of the boundaries, such as for a triangular boundary as in Fig.3**a**, the skin effect is washed out in the corresponding incoherent model, as illustrated in Figs. 3**c-f**. The figures show the spectra under OBC of the Hamiltonian *H* (panel **c**) and of the Markov matrix *M* (panel **e**), along with the spatial distributions of all eigenstates $W(x,y) = (1/N) \sum_l |\psi^{(l)}(x,y)|^2$, where the sum is extend over the number $N=L(L+1)/2$ of the normalized right eigenvectors $\psi^{(l)}(x,y)$ of *H* (panel **d**) and *M* (panel **f**) in the triangular-shaped system. Note that, while on average the eigenstates of *M* are uniformly distributed over the entire sites of the triangle (Fig.3**f**), indicating the absence of the NH skin effect, on average the eigenstates of *H* are localized near the diagonal edge of the triangle (Fig.3**d**), a clear signature of the reciprocal NH skin effect. Finally, we mention that our general analysis could be applied to show washing out of the reciprocal skin effect for incoherent dynamics in other 2D models, such as in the model of Ref.[20].

The above examples indicate that the skin effect persists under incoherent (classical) dynamics when it arises from non-reciprocal hopping amplitudes in the Hamiltonian, but vanishes in reciprocal systems. This result can be extended to special forms of reciprocal and non-reciprocal skin effects, such as the critical skin effect [17,36], where OBC eigenenergies and eigenstates of NH lattice systems jump discontinuously across a critical point in the thermodynamic limit and the system displays scale-free localization, i.e. the localization length of skin modes scales with the system size. This case is considered in Sec.S2 of the Supplementary Material.

**Incoherent skin effect in discrete-time photonic quantum walks.**

Photonic quantum walks have provided a fantastic platform for the observation of a wealth of NH phenomena, such as the non-reciprocal skin effect, NH bulk-boundary correspondence and NH topological phase transitions [19,21,26,50]. Quantum walks also offer feasible systems to introduce controllable decoherence, thus suited to flip from fully coherent to fully incoherent (classical) dynamics [42,43,46]. The spreading laws of the walker in the quantum and classical regimes are well known [53] and, remarkably, quantum coherence yields a faster spreading on the lattice than classical random walks (ballistic versus diffusive). Here we suggest simulating the incoherent dynamics of local excitations along a dissipative lattice displaying the non-reciprocal NH skin effect using photonic quantum walks in well-established protocols, such as those based on a time-multiplexed configuration in fibre networks [19,21,26,49,50]. A schematic of the discrete-time quantum walk is shown in Fig.4**a**. The state vector of the system is defined by

$$|\psi(t)\rangle = \sum_n ( u_n^{(t)} |n\rangle \otimes |H\rangle + v_n^{(t)} |n\rangle \otimes |V\rangle), \qquad (7)$$

where $n$ is the spatial position of the walker on a 1D lattice and $H,V$ denote the internal degree of freedom of the walker (for example the horizontal $H$ or vertical $V$ polarization state of the photon). The variables $u_n^{(t)}$ and $v_n^{(t)}$ are the (non-normalized) amplitude probabilities to find the walker, at discrete time step $t$, at lattice site $n$ and with the internal state $H$ or $V$, respectively. The non-normalized probability $P_n^{(t)}$ to find the walker at lattice site n, regardless of its internal state, is thus $P_n^{(t)} = \left|u_n^{(t)}\right|^2 + \left|v_n^{(t)}\right|^2$. Under coherent dynamics, the state vector evolves according to

$$|\psi(t+1)\rangle = \widehat{U}_{coh} |\psi(t)\rangle \qquad (8)$$

where the one-step propagator $\widehat{U}_{coh}$ is given by the composition of three operations, namely

$$\hat{U}_{coh} = \hat{K}(\gamma)\hat{S}\hat{C}(\theta) \qquad (9)$$

where

$$\hat{S} = \sum_n (|n-1\rangle\langle n| \otimes |H\rangle\langle H| + |n+1\rangle\langle n| \otimes |V\rangle\langle V|) \qquad (10)$$

is the conditional spatial shift operator,

$$\hat{C}(\theta) = \sum_n \begin{pmatrix} \cos\theta & i\sin\theta \\ i\sin\theta & \cos\theta \end{pmatrix} \otimes |n\rangle\langle n| \qquad (11)$$

is the coin operator with rotation angle $\theta$, and

$$\hat{K}(\gamma) = \sum_n \begin{pmatrix} \exp(\gamma) & 0 \\ 0 & \exp(-\gamma) \end{pmatrix} \otimes |n\rangle\langle n| \qquad (12)$$

is the NH operator that introduces an imaginary gauge phase $\gamma$ and is responsible for the appearance of the non-reciprocal skin effect. The coherent evolution for the amplitudes $u_n^{(t)}$ and $v_n^{(t)}$ reads explicitly

$$u_n^{(t+1)} = \exp(\gamma)\left(\cos\theta\, u_{n+1}^{(t)} + i\sin\theta\, v_{n+1}^{(t)}\right) \qquad (13)$$

$$v_n^{(t+1)} = \exp(-\gamma)\left(\cos\theta\, v_{n-1}^{(t)} + i\sin\theta\, u_{n-1}^{(t)}\right) \qquad (14)$$

The Hamiltonian $\hat{H}$ describing the coherent evolution is derived from the relation $\hat{U}_{coh} = \exp(-i\hat{H})$ and in Bloch space its form $H(k)$ is explicitly derived in the Supplementary Material (Sec.S3). The eigenvalues of $H(k)$ are given by $E_\pm(k) = \pm a(k)$, where we have set

$$a(k) \equiv \text{acos}\{\cos\theta \cos(k - i\gamma)\} \qquad (15)$$

and $k$ is the Bloch wave number. Under PBC, $k$ is real and spans the range $-\pi \leq k < \pi$. Correspondingly, the PBC spectrum forms two closed loops in complex energy plane (Fig.4**b**), which is the fingerprint of the NH skin effect. Under OBC, the energy spectrum is entirely real (Fig.4**b**) and is obtained from the same expression $E_\pm(k) = \pm a(k)$ where now $k = q + i\gamma$ is complexified ($-\pi \leq q < \pi$) and varies on the generalized Brillouin zone. In fact, under OBC the NH phase $\gamma$ in Eqs.(13) and (14) can be removed by the non-unitary gauge transformation $u_n^{(t)} \to u_n^{(t)} \exp(-\gamma n)$ and $v_n^{(t)} \to v_n^{(t)} \exp(-\gamma n)$, which in Bloch space is equivalent to complexification of the Bloch wave number [9,11,12]. Such a non-unitary gauge transformation also explains the exponential localization of all eigenstates of $H$ at one edge of the lattice. In bulk dynamics, the NH skin effect can be visualized as a chiral drift dynamics of the walker along the lattice for rather generic initial state of the walker; an example of drift dynamics is shown in Fig.4**c**. The drift velocity can be calculated by standard asymptotic methods and reads (details are given in Sec.S3 of the Supplementary Material)

$$v_{coh} = \pm \frac{\cos\theta \cosh\gamma}{\sqrt{1+\cos^2\theta \sinh^2\gamma}} \tag{16}$$

where the sign depends on the sign of $\gamma$.

The incoherent quantum walk is obtained by multiplying, at each time step, the amplitudes $u_n^{(t)}$ and $v_n^{(t)}$ by random and uncorrelated phase terms [42,43]. This implies that, on average, at each time step the probabilities of the walker to shift on the left or right sides of the lattice, depending on the coin state, sum up incoherently. After letting $X_n^{(t)} = \overline{|u_n^{(t)}|^2}$ and $Y_n^{(t)} = \overline{|v_n^{(t)}|^2}$, where the overline denotes statistical average, the resulting incoherent quantum walk is thus described by the following classical random-walk map

$$X_n^{(t+1)} = \exp(2\gamma)\left(\cos^2\theta\, X_{n+1}^{(t)} + \sin^2\theta\, Y_{n+1}^{(t)}\right) \tag{17}$$

$$Y_n^{(t+1)} = \exp(-2\gamma)\left(\sin^2\theta\, X_{n-1}^{(t)} + \cos^2\theta\, Y_{n-1}^{(t)}\right) \tag{18}$$

which defines the incoherent propagator $\hat{U}_{inc}$ in one time step. The corresponding Markov transition matrix $M$ is obtained from the relation $\hat{U}_{inc} = \exp(\hat{M})$, and its explicit form in Bloch space, $M(k)$, is derived in the Supplementary Material (Sec.S3). The eigenvalues of $M(k)$ are formed by the two branches

$$\lambda_\pm(k) = \frac{1}{2}\ln(2\cos^2\theta - 1) \pm ia'(k) \tag{19}$$

where $k$ is the Bloch wave number and where we have set

$$a'(k) = \text{acos}\left\{\frac{\cos^2\theta}{\sqrt{2\cos^2\theta - 1}}\cos(k - 2i\gamma)\right\} \tag{20}$$

Under PBC, $k$ is real and varies in the range $-\pi \leq k \leq \pi$; correspondingly the eigenvalues of the Markov matrix describe two closed loops in complex plane (Fig.4**d**). On the other hand, under OBC is complexified and takes the values $k = q + 2i\gamma$, with $-\pi \leq q \leq \pi$; the corresponding energy spectrum describes open segments (Fig.4**d**), and the eigenstates are squeezed toward one edge of the lattice, corresponding to the incoherent NH skin effect. We note that for the special case of Hadamard coin $\theta = \pi/4$ the elements and eigenvalues of the Markov transition matrix in Bloch space apparently diverge [Eqs.(19) and (20)], owing to the fact that the incoherent one-step propagator matrix $U_{inc}(k)$ displays a vanishing eigenvalue (more technical details are given in Sec.S3 of the Supplementary Material). However, such a divergence is not of physical relevance and the incoherent quantum walk dynamics for the Hadamard coin $\theta = \pi/4$ can be solved analytically in a simple way (Sec.S4 of the Supplementary Material), the (unnormalized) occupation probability $P_n^{(t)} = X_n^{(t)} + Y_n^{(t)}$ being given by an asymmetric binomial distribution.

In bulk dynamics, the skin effect is visualized as a unidirectional drift of any initial state of the walker; an example of drift dynamics is shown in Fig.4**e**. The drift velocity for the incoherent quantum walk can be calculated from an asymptotic analysis and reads (Sec.S3 of the Supplementary Material)

$$v_{inc} = \frac{\cos^2\theta \sinh(2\gamma)}{\sqrt{\sin^4\theta + \cos^4\theta \sinh^2(2\gamma)}}. \qquad (21)$$

It is worth commenting on the behavior of the drift velocities $v_{coh}$ and $v_{inc}$, for the coherent and incoherent quantum walks, as the NH gauge phase $\gamma$ is increased above zero. A typical behavior of the two velocities versus $\gamma$, for a coin angle close to the Hadamard coin, is shown in Fig.4**f**. Note that, while for coherent dynamics the drift velocity is non-vanishing for any arbitrarily small value of $\gamma$ [Eq.(16)], in the incoherent quantum walk the drift velocity vanishes as $\gamma \to 0$ [Eq.(21)]. This behavior can be explained by the different spreading dynamics of quantum and classical random walks in the Hermitian limit $\gamma = 0$ [53]. In a quantum walk the spreading is ballistic and the excitation asymptotically spreads with two main peaks centered along the two space-time lines $n = \pm vt$, where $v = \cos\theta$ is precisely the limit of $v_{coh}$ as $\gamma \to 0$. As $\gamma$ is slightly increased above zero, one of the two peaks is amplified while the other one is attenuated and asymptotically dies, resulting in an irreversible drift along the dominant peak (Fig.S2 in the Supplementary Material). Conversely, in the classical random walk the spreading is diffusive and there is one main peak along the space-time line $n = 0$. In this case, as $\gamma$ is slightly increased above zero, the drift velocity remains small. However, a $\gamma$ is further increased, the drift velocity in the incoherent quantum walk can overcome the drift velocity of the coherent regime (Fig.4**f**). This means that, very interestingly, in NH quantum walks dephasing effects, leading to classicalization of the dynamics, can enhance excitation transport in the lattice, contrary to what happens in the Hermitian case where transport is always faster under quantum coherence.

## Discussion

The NH skin effect, i.e. the phenomenon that eigenstates of a NH Hamiltonian mainly reside at the boundary of the system rather than in the bulk, provides one of the most exotic manifestations of point-gap topology in non-Hermitian systems. Understanding and exploiting the NH skin effect can have practical implications for designing devices with unique properties, especially in the context of photonics and quantum technologies. For example, in photonics the NH skin effect may be harnessed for light funneling [21], for guiding light in robust ways [54] and for designing novel kinds of topological lasers [55,56]. So far, the NH skin effect has been unravelled in systems displaying full or partial wave coherence, while the fate of the NH skin effect when coherence is lost – a regime which is commonplace in many complex physical, chemical and biological systems out of equilibrium– remains largely unexplored. Here we have investigated the fate of the NH skin effect in the fully incoherent regime, showing that the effect persists under incoherent dynamics when it originates from non-reciprocal hopping in the system, while reciprocal skin effect is washed out by dephasing. The results have been illustrated by considering incoherent photonic quantum walks in synthetic lattices, which should provide an experimentally accessible platform for the observation of incoherent non-reciprocal skin effect. Interestingly, while in Hermitian quantum walks decoherence leads to transport slowing down, in NH quantum walks dephasing effects can make transport faster. The present study provides major advancements in the understanding of NH skin effect and is expected to stimulate further studies on an emergent and impactful area of research.

## Materials and Methods

The numerical simulations of wave packet evolution in the bulk of the lattice under incoherent dynamics is obtained by propagating the initial state of the system $|\psi(0)\rangle$ along a large number $S$ of trajectories (Fig.1**b**), and then making the statistical average of the unnormalized occupation probabilities $P_n(t = N\Delta t) = \overline{|\langle n|\psi(t)\rangle|^2}$. In each trajectory the state vector of the system evolves according to $|\psi(N\Delta t)\rangle = \prod_{\alpha=1}^{N} \hat{P}_\alpha \hat{U}_{coh}(\Delta t) |\psi(0)\rangle$, where $\hat{U}_{coh}(\Delta t) = \exp(-i\hat{H}\Delta t)$ describes the

coherent propagation of the system for the time interval $\Delta t$ ($\Delta t=1$ for the quantum walk model), $\hat{P}_\alpha = \sum_n \exp(i\phi_n^{(\alpha)}) |n\rangle\langle n|$ describes the stochastic phase shift operation, and $\phi_n^{(\alpha)}$ are the uncorrelated random phases applied at lattice site $n$ and time step $\alpha$, uniformly distributed in the range $(-\pi, \pi)$. The propagation is performed in physical space on a wide enough lattice size to avoid edge effects at the largest propagation time. The coherent propagator $\hat{U}_{coh}(\Delta t)$ has been computed in physical space by using exponential matrix function in MatLab. In the snapshots of wave packet evolution shown in Figs.2 and 4, after each time step $\Delta t$ the wave function $|\psi(t)\rangle$ has been renormalized as $|\psi(t)\rangle \to |\psi(t)\rangle/\||\psi(t)\rangle\|$, where $\||\psi(t)\rangle\|$ is the norm of the wave function.

## Data availability

Any related theoretical or numerical information not mentioned in the text and other findings of this study are available from the corresponding author upon reasonable request.

## Acknowledgments

This work has been supported by the Spanish State Research Agency, through the Severo Ochoa and Maria de Maeztu Program for Centers and Units of Excellence in R&D (Grant No. MDM-2017-0711).

## Competing interest declaration

The author declares no competing interests.

## Additional information

**Supplementary information** accompanies this paper at XXXXXX

Correspondence and requests for materials should be addressed to Stefano Longhi (stefano.longhi@polimi.it)

# Figures and captions

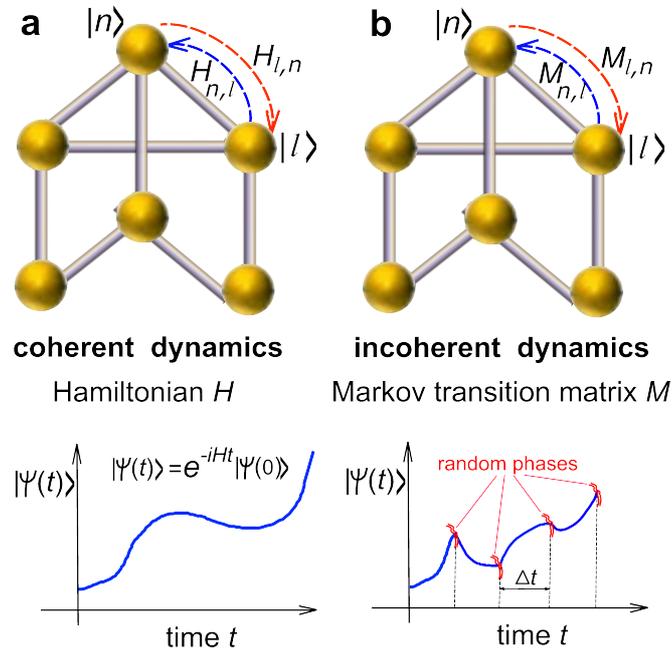

**coherent dynamics**  
Hamiltonian $H$

**incoherent dynamics**  
Markov transition matrix $M$

**Fig. 1.** **Model. a** Schematic of a lattice system (network) comprising $N$ sites (nodes). Under coherent dynamics the system is described by a NH tight-binding matrix Hamiltonian $H_{n,l}$ and the state vector $|\psi(t)\rangle$ evolves in time along a single trajectory. **b** Under incoherent dynamics, with randomized phases in each lattice site at successive time intervals $\Delta t$, the trajectory of the state vector is modified. After statistical averaging over all different trajectories, corresponding to different realizations of stochastic phases, excitation transfer among the nodes of the network is described by a Markov transition matrix $M_{n,l}$ (classical random walk). The relation between the matrices $H$ and $M$ is given by Eq.(3).

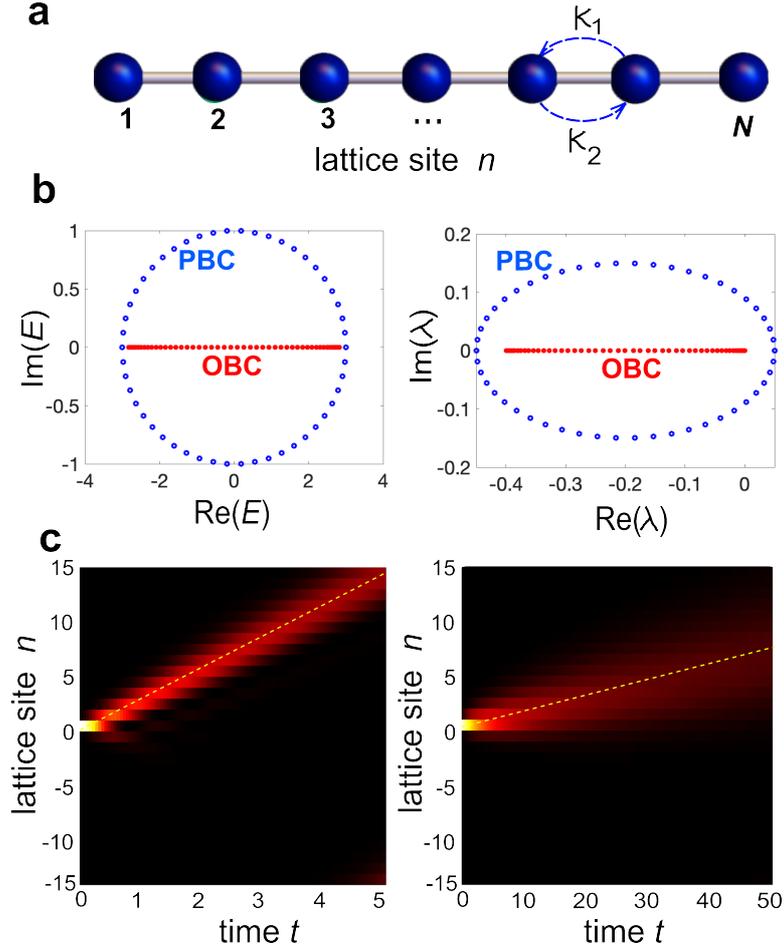

**Fig. 2. Non-reciprocal skin effect**. a Schematic of the 1D lattice with asymmetric hopping amplitudes $\kappa_1$ and $\kappa_2$ (Hatano-Nelson model). b Left panel: energy spectrum $E$ in complex plane of the Hatano-Nelson Hamiltonian $H$ under PBC and OBC for parameter values $\kappa_1=1$ and $\kappa_2=2$. Lattice size $N=50$. Right panel: corresponding spectrum $\lambda$ of the Markov transition matrix $M$ for $\Delta t = 0.05$. c Bulk dynamics (snapshot of $|\psi_n(t)|^2$, normalized at each time step to its norm, on a pseudocolor map) under coherent (left panel) and incoherent (right panel) regimes for initial single-site excitation of the lattice. In the incoherent regime, the distribution $|\psi_n(t)|^2$ has been obtained after averaging over $S=1000$ trajectories corresponding to different realizations of the stochastic phases (see Materials and Methods). The dashed straight lines indicate the asymptotic drift dynamics at the speeds $v_{coh} = (\kappa_1 + \kappa_2)$ and $v_{inc} = \Delta t(\kappa_2^2 - \kappa_1^2)$ (right panel).

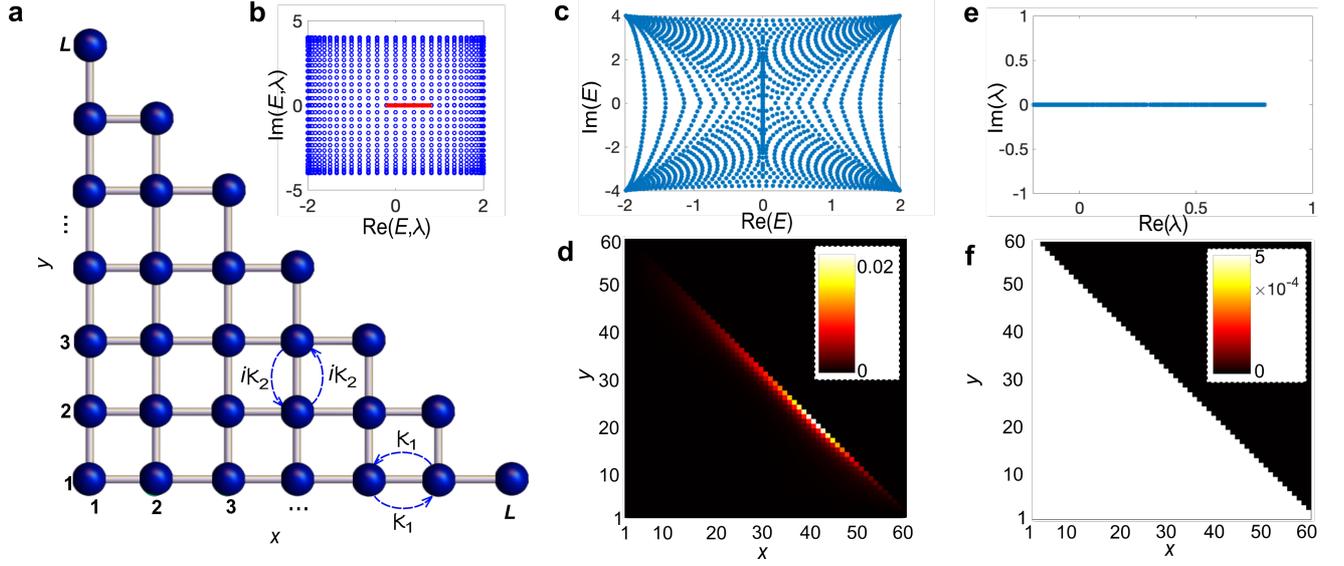

**Fig. 3. Reciprocal skin effect in 2D lattices**. **a** Schematic of a 2D square lattice with reciprocal hopping amplitudes $\kappa_1$ in the horizontal $x$ direction and $i\kappa_2$ in the vertical $y$ direction. OBC are assumed along a triangular contour. The total number of sites in the lattice with OBC is $N=L(L+1)/2$. For PBC, a square lattice of size $L \times L$ is instead assumed. **b** PBC spectra of the Hamiltonian $H$ (open blue circles) and of the Markov matrix $M$ (filled red circles) in complex plane. System size is $L$=60, hopping amplitudes are $\kappa_1$=1 and $i\kappa_2 = 2i$. **c** OBC energy spectrum of the Hamiltonian $H$ for the triangular contour of panel **a** ($L$=60). **d** Spatial distribution of all eigenstates $W(x,y)$ of $H$ in the triangular-shaped system. **e-f** same as panels **c-d**, but for the Markov matrix $M$ with $\Delta t = 0.05$.

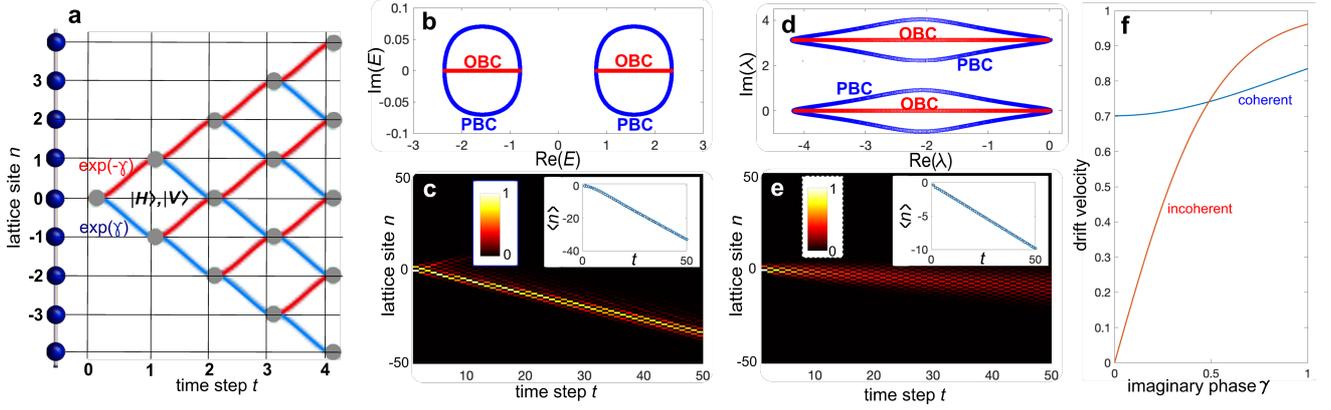

**Fig. 4. Coherent vs incoherent photonic quantum walks**. **a** Schematic of a 1D quantum walk. *H* and *V* are the two internal degrees of freedom of the walker. **b** Energy spectrum of the Hamiltonian *H*, for either PBC or OBC, under coherent dynamics. Parameter values are $\beta = 1.01 \times \pi/4$ and $\gamma = 0.1$. **c** Wave dynamics of the quantum walk for coherent dynamics (snapshot of normalized occupation probability at various lattice sites *n* versus time step *t*). The system is initially prepared in the state $|\psi(0)\rangle = (1/\sqrt{2})(|H\rangle + |V\rangle) \otimes |0\rangle$. The inset shows the behavior of the wave packet center of mass $\langle n \rangle$ (open circles); the dashed curve, almost overlapped with circles, describes the uniform drift motion at the speed predicted by the asymptotic analysis. **d-e** Same as **b-c**, but for the incoherent dynamics. Panel **d** is the spectrum $\lambda$ of the Markov transfer matrix *M*. The normalized evolution of probability distribution in panel **e** is obtained from the statistical average of *S*=1000 trajectories corresponding to different realizations of stochastic phases (see Materials and Methods). **f** Behavior of the drift velocity versus the imaginary gauge phase $\gamma$ for the coherent ($v_{coh}$) and incoherent ($v_{inc}$) quantum walks. The coin angle is $\theta = 1.01 \times \pi/4$.